# Golden, Quasicrystalline, Chiral Packings of Tetrahedra

Fang Fang[a], Garrett Sadler[a], Julio Kovacs, Klee Irwin[a, b]
*Quantum Gravity Research Group, Topanga, California 90290, USA*

**Since antiquity, the packing of convex shapes has been of great interest to many scientists and mathematicians [1-7]. Recently, particular interest has been given to packings of three-dimensional tetrahedra [8-20]. Dense packings of both crystalline [8, 10, 15, 17, 19] and semi-quasicrystalline [14] have been reported. It is interesting that a semi-quasicrystalline packing of tetrahedra can emerge naturally within a thermodynamic simulation approach [14]. However, this packing is not perfectly quasicrystalline and the packing density, while dense, is not maximal. Here we suggest that a "golden rotation" between tetrahedral facial junctions can arrange tetrahedra into a perfect quasicrystalline packing. Using this golden rotation, tetrahedra can be organized into "triangular", "pentagonal", and "spherical" locally dense aggregates. Additionally, the aperiodic Boerdijk-Coxeter helix [23, 24] (tetrahelix) is transformed into a structure of 3- or 5-fold periodicity—depending on the relative chiralities of the helix and rotation—herein referred to as the "philix". Further, using this same rotation, we build (1) a shell structure which resembles a Penrose tiling upon projection into two dimensions, and (2) a "tetragrid" structure assembled of golden rhombohedral unit cells. Our results indicate that this rotation is closely associated with Fuller's "jitterbug transformation" [21] and that the total number of face-plane classes (defined below) is significantly reduced in comparison with general tetrahedral aggregations, suggesting a quasicrystalline packing of tetrahedra which is both dynamic and dense. The golden rotation that we report presents a novel tool for arranging tetrahedra into perfect quasicrystalline, dense packings.**

Packings of spheres and the platonic solids have been of great interest to mathematicians since ancient times [1-7]. With in the past few years, rapid progress has been made in the problem of dense packings of tetrahedra [9-20]. Previous studies have mainly focus on crystalline packings [8, 10, 15, 17, 19] and have since appeared to reach a plateau. Recently, the possibility of quasicrystals displaying tetrahedral phases has been examined [27], and a subsequent report [14] has interestingly and unexpectedly produced a quasicrystalline packing of tetrahedra, motivating the present work. In fact, the close relationship between the tetrahedron, icosahedron, and pentagonal bipyramid (discussed below) strongly suggests the existence of a dense, quasicrystalline packing of tetrahedra, as pentagonal and icosahedral symmetry are forbidden in crystals but are common in quasicrystals (QC) [22].

In our analysis of tetrahedral packings and assemblages, we have found utility in a geometric object's "plane classes". Planes are said to belong to the same plane class if and only if their normal vectors are parallel. The number of plane classes for a collection of polytopes is defined as the number of distinct plane classes comprising the collection's two-dimensional faces. A characteristic feature of crystals and QCs is that they have a finite number of plane classes. For example, a three-dimensional cubic lattice has a total of three plane classes. A three-dimensional

---
a These authors contributed equally to this work.
b Group director. Correspondence should be directed to: klee@quantumgravityresearch.org

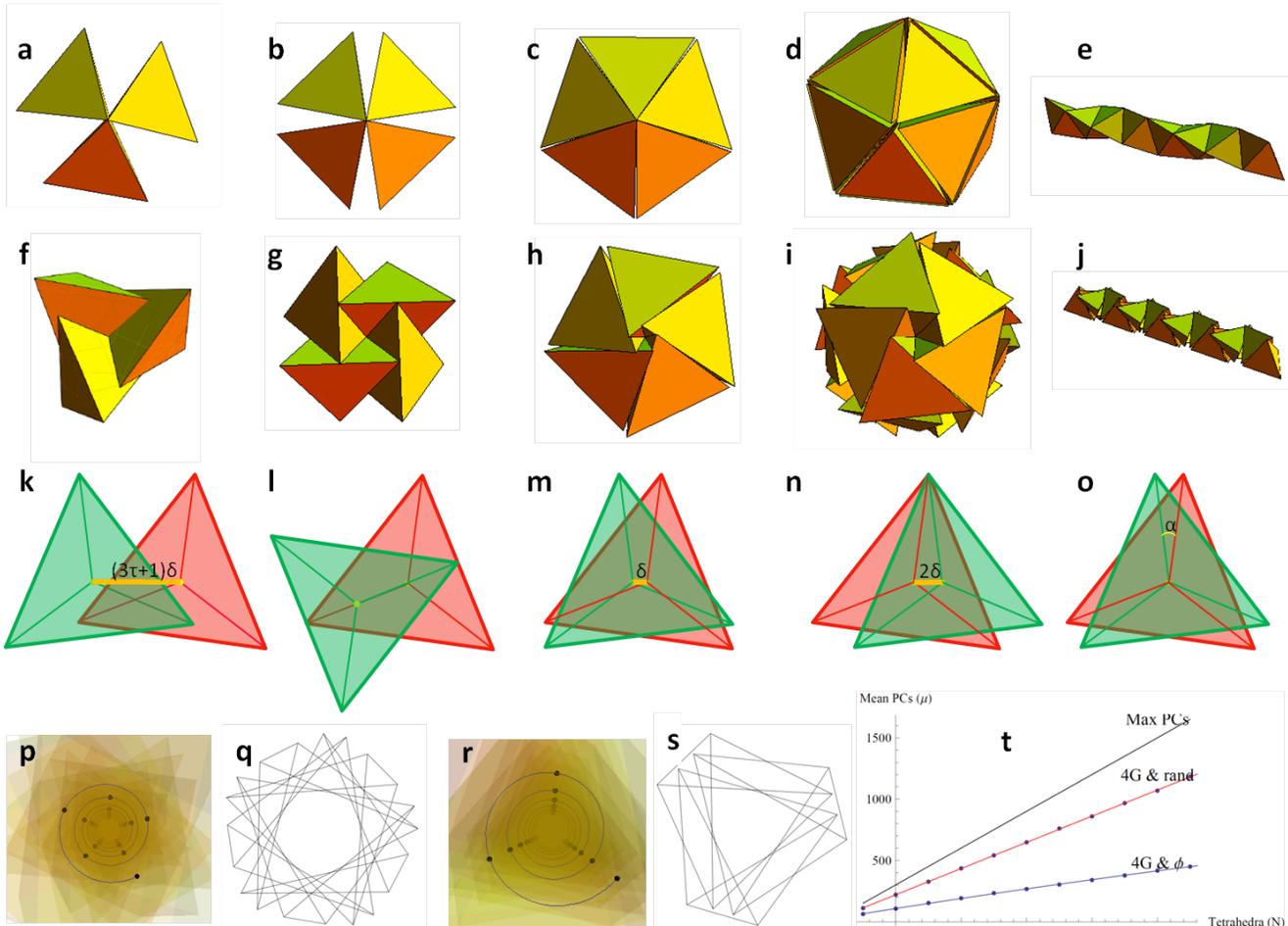

**Figure 1 | Tetrahedral structures subjected to the golden rotation and associated junction types.** Groups of tetrahedra uniformly distributed about a common edge (a–c), and vertex (d). A linear arrangement of tetrahedra forms the Boerdijk-Coxeter helix (tetrahelix) (e). "Twisting" these structures eliminates inter-tetrahedral gaps and reduces the number of plane classes (f–j). The corresponding projections of facial junctions are shown in (k–o). The relative angular relationship between faces in k, m, n, and o is related to the golden ratio. Hence, this rotational relationship is referred to as the golden rotation. In its canonical form, the Boerdijk-Coxeter helix is aperiodic, however, following application of the golden rotation, this structure forms the "philix" (j) with a 3- or 5-tetrahedron periodicity, depending on the relative chiralities of the helix and rotation. Axial projections of this helix show the 5-period (p, q) and 3-period (r, s) cases. The significant reduction of plane classes obtained using these face junctions vs. a randomly generated set is shown in (t).

generalization of the Penrose tiling [28, 29] may be put into correspondence with rhombic triacontahedron, and consequently comprises 15 plane classes. Finally, the tetrahedral, quasicrystalline approximant produced by Haji-Akbari *et al.* [14], in principle, comprises an infinite number of plane classes due to the underlying stochastic nature of that structure's effecting Monte Carlo simulation.

The irrationally valued dihedral angle of the tetrahedron presents a difficulty towards the production of tetrahedral packings with long-range order and/or a finite number of plane classes. As an extreme example, consider the Boerdijk-Coxeter helix [23, 24] depicted in Figure 1e. This structure is well-known to be aperiodic and, if extended to comprise an infinite number of tetrahedra, produces infinitely many plane classes.

In the pursuit of a tetrahedral structure with long-range order and a small number of plane classes, we have found useful five categories of face junctions—defined as orientational facial relationships between coincident tetrahedra. These face junctions are shown in Figures 1k–o and are referred to here as 3G, 4G, 5G, 20G, and FC (face centered), respectively. The 3G, 4G, and 5G face junctions may be obtained by "twisting" the $n$ tetrahedra of a collection ($n < 6$) arranged about a common edge (Figures 1a–c) by an angle $\alpha_n$, expressed as

$$\alpha_n = \tan^{-1}\left(\frac{\sqrt{\cos^2\left(\frac{\theta}{2}\right) - \cos^2\left(\frac{\theta_n}{2}\right)}}{\sin\left(\frac{\theta}{2}\right)\cos\left(\frac{\theta_n}{2}\right)}\right) \quad (1)$$

where $\theta = \arccos(1/3)$ is the tetrahedral dihedral angle and $\theta_n = 2\pi/n$, about an axis passing between the midpoints of the central and peripheral edge. (In the case of the 20G, tetrahedra are rotated by an angle of $\arccos((\sqrt{5}+3)/(4\sqrt{2}))$ about an axis that passes from the icosahedral center through each tetrahedron's exposed face.) This procedure is equivalent to twisting the tetrahedra of Figures 1a–d uniformly until all gaps have been closed (Figures 1f–i). The resulting angular relationship between faces in a junction pair is determined by

$$\beta_n = 2\tan^{-1}\left(\frac{\sqrt{\cos^2\left(\frac{\theta}{2}\right) - \cos^2\left(\frac{\theta_n}{2}\right)}}{\cos\left(\frac{\theta_n}{2}\right)}\right) \quad (2)$$

| Type | Twisting angle, $\alpha_n$ | Facial rotation angle, $\beta_n$ | Facial center offset | Symmetry | Num. plane classes |
|---|---|---|---|---|---|
| 3G | $\arccos\left(\frac{1}{\sqrt{6}}\right) \approx 65.91°$ | $\arccos\left(-\frac{1}{4}\right) \approx 104.48°$ | $\frac{3\tau+1}{2\sqrt{6}\,\tau^2} \approx 0.4564$ | 3-fold | 9 |
| 4G | $\arccos\left(\frac{1}{\sqrt{2}}\right) = 45°$ | $\arccos\left(\frac{1}{2}\right) = 60°$ | $\frac{1}{2\sqrt{3}} \approx 0.29$ | 4-fold | 4 |
| 5G | $\arccos\left(\sqrt{\frac{1}{2}+\frac{1}{\sqrt{5}}}\right)$ $= \arccos\left(\frac{\tau^2}{\sqrt{2(\tau+2)}}\right) \approx 13.28°$ | $\beta_g \equiv \arccos\left(\frac{1+3\sqrt{5}}{8}\right)$ $= \arccos\left(\frac{3\tau-1}{4}\right) \approx 15.52°$ | $\frac{1}{\sqrt{6}(\sqrt{5}+3)} = \frac{1}{2\sqrt{6}\,\tau^2} \approx 0.078$ | 5-fold | 10 |
| 20G | $\arccos\left(\frac{\sqrt{5}+3}{4\sqrt{2}}\right)$ $= \arccos\left(\frac{\tau^2}{2\sqrt{2}}\right) \approx 22.24°$ | $\beta_g \equiv \arccos\left(\frac{1+3\sqrt{5}}{8}\right)$ $= \arccos\left(\frac{3\tau-1}{4}\right) \approx 15.52°$ | $\frac{2}{\sqrt{6}(\sqrt{5}+3)} = \frac{1}{\sqrt{6}\,\tau^2} \approx 0.15$ | $A_5$ | 10 |
| FC | n/a | $\beta_g \equiv \arccos\left(\frac{1+3\sqrt{5}}{8}\right)$ $= \arccos\left(\frac{3\tau-1}{4}\right) \approx 15.52°$ | 0 | 3-fold | 9 |
| | | | | 5-fold | 10 |

**Table 1 | Important parameters for facial junction types.** "Twisting" angles are given by $\alpha_n$. The relative angular relationship between face pairs in a junction are given by $\beta_n$. The golden rotation angle is denoted by $\beta_n$. Values for facial center offsets represent the distance between facial centers of a junction pair. Here, $A_5$ denotes the orientation-preserving subgroup of the icosahedral symmetry group, the alternating group on 5 letters.

Assembling tetrahedral structures with these facial junctions has the effect of reducing the total number of plane classes on certain

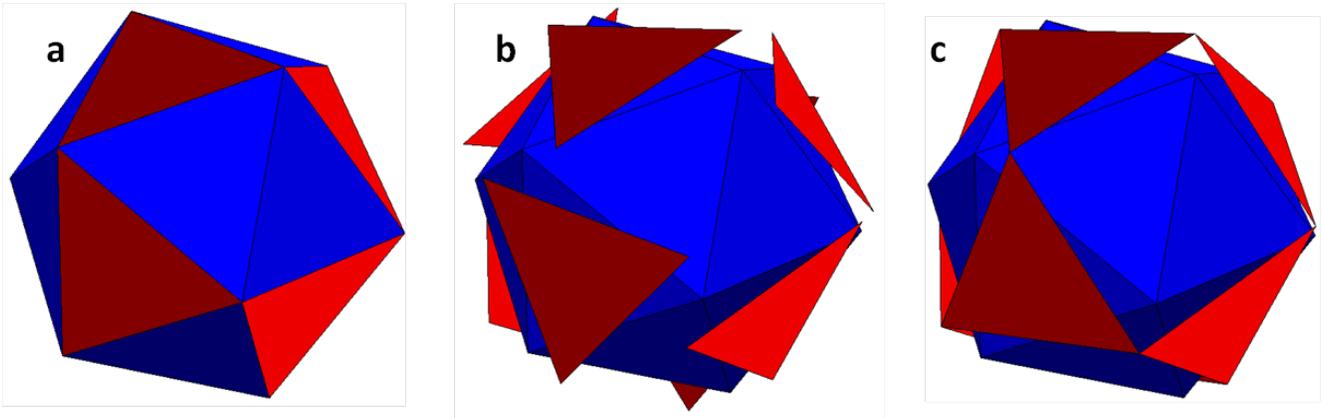

**Figure 2 | Jitterbug transformation associated with handedness of the golden rotation.** 8 selected faces (red) on the icosahedron with octahedral symmetry (a). When rotated through an angle of $\alpha_{20}$ these faces coincide with 8 outer faces of the "twisted" 20G (b). When the handedness of rotation is alternated, these faces coincide with 8 faces of the cuboctohedron (c).

arrangements of tetrahedra (see Figure 1t).

A list of relevant parameters for these "closed-twisted" structures is provided in Table 1. Of note is the fact that the relative face-to-face rotations for the 5G, 20G, and FC junctions are all equal ($\beta_g = \arccos((3\tau - 1)/4) \approx 15.5225°$, where $\tau$ is the golden proportion). Hence, this value is referred to as the golden rotation, $\beta_g$. In these structures—as for the twisted 4G arrangement—the number of plane classes is reduced compared with the uniform arrangements. Additionally, the twisting angle of the 20G is equal to the rotational angle of Fuller's "jitterbug transformation" which transmutes the icosahedron to the cuboctahedron [21]. Taken together, these two features strongly suggest a "transformable" quasicrystalline packing of tetrahedra. (See Figure 2.)

In attempting to construct a dense quasicrystalline packing of tetrahedra, we have employed both bottom-up and top-down approaches. The former consists of using the various face junctions of Figures 1k–o either (1) to construct a skeleton of the quasicrystal, subsequently filling gaps with appropriately shaped tiles (e.g., "the shell model", Figure 3a), or (2) to build construct a primitive cell of the quasicrystal (e.g., the "tetragrid model", Figures 3d,e). These models may be assembled using the following procedures.

**Shell model construction**

1. To an initial tetrahedron, four tetrahedra are appended such that the facial centers of their incident faces coincide. Each appended tetrahedron is then rotated such that the relative angle between the faces is the golden rotation angle, $\beta_g$.

2. Tetrahedra are then iteratively appended to each "exposed" face (i.e., not already affixed to another tetrahedron), and oriented in the same manner as above.

3. Following each iteration, tetrahedra are scaled down (with centers fixed) until all collisions are eliminated.

**Tetragrid model construction**

1. Apply the golden rotation to the constituent tetrahedra of a Boerdijk-Coxeter helix to obtain a "philix" with linear periods of 3 or 5 tetrahedra.

2. The periodic nature of the philix along its axis allows for the formation of three-dimensional grids of philices, by spacing their points of intersection every 3 or 5

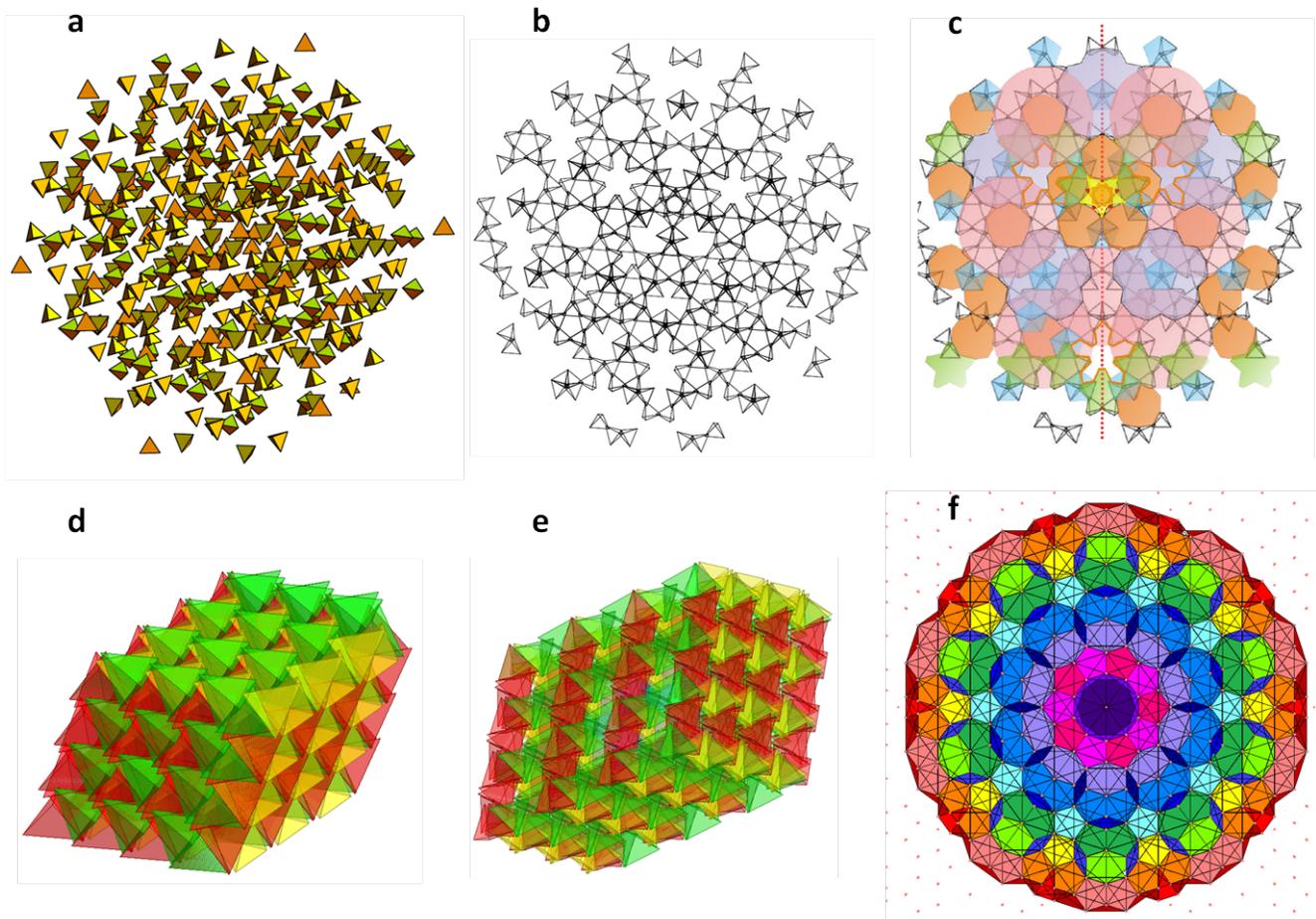

**Figure 3 | The shell and tetragrid models.** 3D shell structure with 5 iterations and a total of 284 tetrahedra (a) and its 2D projection down a 5-fold symmetry axis (b). Coloring this projection reveals similarities to the Penrose tiling (c). Tetragrid structure comprised of 5-periodic (d) and 3-periodic (e) philices. A layer of the 3D projection of *V'*, viewed from an axis of 10-fold symmetry, is shown in (f).

tetrahedra.

3. At each crossing point, there are three directions along which to place philices, each given by one of the opposing faces in a dipyramid (two consecutive tetrahedra sharing a face)

4. Since these grid lines form three families of parallel lines, rhombohedra may be formed. In the case of 3-periodic philices, golden rhombohedra are produced, whereas rhombohedra with a diagonal to edge ratio of $\sqrt{2}$ are formed for 5-periodic philices.

These models are shown in Figures 3a, d, and e. The two-dimensional projection of the shell model very closely resembles a Penrose tiling (Figures 3b, c). In the tetragrid case employing 3-periodic philices, a golden rhombohedral primitive cell is obtained, reminiscent of the golden rhombohedron prototile of the three-dimensional Penrose tiling [28, 29]. These are encouraging indications for the feasibility of generating quasicrystalline packings using the face junctions described above.

Our top-down approach involves the production of a four-dimensional quasicrystal *via*

the cut-and-project method [22, 26] from the $E_8$ lattice. Selected points of this structure are subsequently projected into three-dimensional space. The details of this procedure are as follows:

1. Determine the set of edge of minimal length *l* among all points of the projected (8D to 4D) set. Call this set *E*.

2. From this set, select those vertices that belong to 12 or more of these edges (i.e., vertices forming connections with at least 12 others at a distance of *l*). Let the set of these vertices be denoted by *V*.

3. Select those edges of *E* whose end points are contained in *V*. Denote this set by *E'*, and let *V'* denote the set of endpoints of edges in *E'*.

4. Finally, project the vertices of *V'* from four-dimensions to three-dimensions.

We expect this procedure to produce a three-dimensional quasicrystal of irregular tetrahedra. Through the application of our golden and 4G face junctions, we anticipate the ability to regularize tetrahedral cells while introducing inter-tetrahedral gaps. We have achieved some preliminary results in this direction. Figure 3f shows a layer of the 3D projection from *V'*, viewed along a five-fold axis of symmetry. This layer is clearly a cartwheel-type quasicrystal, similar to the two-dimensional projection of the shell model. An interesting fact about this projection method is its relationship with the "Sum of Squares" law [25], which states that for any edge-transitive polytope projected orthographically from *m* dimensions to *n* dimensions, the ratio of the sum of squared edge lengths is $m/n$ ($m > n > 0$).

In conclusion, the golden rotation described in this writing produces plane-class reduction in tetrahedral packings, and is closely related to Fuller's jitterbug transformation. Consequently, we suspect that a tetrahedral quasicrystal may be constructed using this golden rotation, and that a dynamic quasicrystal may be achieved using jitterbug-like transformation between consecutive quasicrystal frames. Towards this goal, both bottom-up and top-down approaches are currently under investigation.